# Revisiting sample size planning for receiver operating characteristic studies: a confidence interval approach with precision and assurance

Di Shu[1,2,3] and Guangyong Zou[4,5]

[1]Department of Biostatistics, Epidemiology and Informatics, University of Pennsylvania Perelman School of Medicine, Philadelphia, Pennsylvania, USA

[2]Department of Pediatrics, University of Pennsylvania Perelman School of Medicine, Philadelphia, Pennsylvania, USA

[3]Center for Pediatric Clinical Effectiveness, Children's Hospital of Philadelphia, Philadelphia, Pennsylvania, USA

[4]Department of Epidemiology and Biostatistics, Schulich School of Medicine & Dentistry, Western University, London, Ontario, Canada

[5]Robarts Research Institute, Western University, London, Ontario, Canada

Correspondence

Di Shu, Department of Biostatistics, Epidemiology and Informatics, University of Pennsylvania Perelman School of Medicine, 423 Guardian Drive, Philadelphia, PA 19104-6021, USA. Office Tel: 215-573-4887. Email: di.shu@pennmedicine.upenn.edu

**Abstract**

**Objectives:** Estimation of areas under receiver operating characteristic curves (AUCs) and their differences is a key task in diagnostic studies. We aimed to derive, evaluate, and implement simple sample size formulas for such studies with a focus on estimation rather than hypothesis testing.

**Materials and Methods:** Sample size formulas were developed by explicitly incorporating pre-specified precision and assurance, with precision denoted by the lower limit of confidence interval and assurance denoted by the probability of achieving that lower limit. A new variance function was proposed for valid estimation allowing for unequal variances of observations in the disease and non-disease groups. Performance of the proposed formulas was evaluated through simulation.

**Results:** Closed-form sample size formulas were obtained. Simulation results demonstrated that the proposed formulas produced empirical assurance probability close to the pre-specified assurance probability and empirical coverage probability close to the nominal 95%. Real-world worked examples were presented for illustration.

**Conclusions:** Sample size formulas based on estimation of AUCs and their differences were developed. Simulation results suggested good performance in terms of achieving pre-specified precision and assurance probability. An online calculator for implementing the proposed formulas is openly available at https://dishu.page/calculator/.

# INTRODUCTION

In diagnostic studies with data of ordinal or continuous scale, the AUC, i.e., area under the receiver operating characteristic (ROC) curve, is commonly used to quantify the overall accuracy of the diagnostic tool (commonly referred to as test).[1-3] The ROC curve is a plot of sensitivity against 1-specificity obtained by varying the cutoff point above which a participant will be considered test positive (diseased). The AUC represents the probability of correctly identifying the participant condition from a random pair of diseased and non-diseased participants.[1,4] The values of AUC typically range from 0.5 (no apparent ability) to 1 (perfect accuracy). Comparison of two diagnostic tools can be accomplished via inference on the difference between two AUCs.[5]

Sample size estimation is the first task in planning of an ROC study.[6] Inadequate and overly-planned sample sizes both can considerably waste research resources; underpowered studies might even be considered unethical.[7] Previous sample size formulas and software packages were primarily developed for achieving either a desired power in hypothesis testing or a desired expected confidence interval width.[1,5,8-16] Here we focus on a different criterion, whether the estimates are of acceptable precision, as quantified by the lower confidence limit.[17-19] Note that hypothesis testing focuses on a single value (the null hypothesis) while confidence interval estimation focuses on a range of plausible values and thus is more informative than the former approach.[20] Based on available benchmarks[21, 22] — for example for certain diagnostic tools an AUC between 0.9 to 1.0 reflects 'excellent' accuracy, 0.8 to 0.9 'good', 0.7 to 0.8 'fair', 0.6 to 0.7 'poor', and below 0.6 'failed' — one might ask "how many participants should be recruited to assure with 90% chance that the lower confidence limit is above 0.8 (so that I feel comfortable to call the test good)?"

The purpose of this report is to derive, evaluate, and implement simple sample size formulas for planning ROC studies in order to achieve acceptable precision in the estimation of single AUCs and their difference. Our approach aligns well with the trend towards using confidence intervals for inferences,[20,23] by incorporating a pre-specified probability, which is referred to as assurance, of achieving the desired lower confidence limit.

## MATERIALS AND METHODS

**Proposed sample size formula: estimation of a single AUC**

We first derive a sample size formula for achieving a pre-specified lower bound of the two-sided 95% confidence interval for the AUC with assurance probability $1 - \beta$. Let $\theta$ and $\theta_0$ denote the pre-specified AUC and the lower bound, respectively. Here $1 - \beta$ is called the assurance probability because we aim to compute required sample size such that $P(\hat{\theta}_L \geq \theta_0) \geq 1 - \beta$ for a given $\theta > \theta_0$, where $\hat{\theta}_L$ denotes the lower confidence limit for $\theta$. One may regard the assurance here as equivalent to power in hypothesis testing that focuses on a specific null value, but presenting as assurance probability is more coherent with the goal of estimation, which is to find the range of plausible values based on the data at hand.

Specifically, consider testing a null hypothesis $H_0: \text{AUC} = \theta_0$ against an alternative hypothesis $H_1: \text{AUC} = \theta > \theta_0$ with a rejection condition $\hat{\theta}_L \geq \theta_0$. This hypothesis test is closely related to the assurance approach because its power is $1 - \beta$. However, it should be noted that the resulting level of test, $P(\hat{\theta}_L \geq \theta_0 \mid \text{AUC} = \theta_0)$, generally does not equal to $\alpha$, the significance level of the confidence interval and the target level for a one-tail test. For example, when $\sqrt{n}(\hat{\theta} - \theta_0)$ exactly follows a normal distribution $N(0, \sigma^2)$ and that the confidence interval is symmetric, the test level $P(\hat{\theta} - z_{\alpha/2}\hat{\sigma}/\sqrt{n} \geq \theta_0 \mid \text{AUC} = \theta_0) = \alpha/2 < \alpha$, where $\hat{\sigma}$ is an estimate of $\sigma$ and $z_{\alpha/2}$ is the upper $\alpha/2$ quantile of the standard normal distribution. Such test provides a less transparent interpretation of sample size estimation than the assurance probability.

Prior studies noted poor coverage from Wald-type intervals when sample size is relatively small.[2,24-26] Therefore, we adopt asymmetric confidence intervals based on the logit transform which were found to have better coverage.[2,24-26] We also assume data from both groups are normally distributed.[8,9]

With detailed derivations presented in the Online Supplementary Material, we propose the following sample size formula:

$$n = \left\{\frac{z_\beta + z_{\alpha/2}}{\text{logit}(\theta) - \text{logit}(\theta_0)}\right\}^2 \frac{f(\theta)}{\theta^2(1-\theta)^2} \frac{\pi}{3}$$

where the logit transformation is defined as $\text{logit}(\theta) = \log\left(\frac{\theta}{1-\theta}\right)$, $z_\beta$ is the upper $\beta$ quantile of the standard normal distribution, $1 - \alpha$ is the confidence level (conventionally, $\alpha = 0.05$), and $f(\theta)$ is the kernel of the variance for the estimated AUC (factoring out the sample size), that is, $f(\theta)/n = \text{var}(\hat{\theta})$. When $\beta = 0.5$, the sample size reduces to that for achieving an average (or expected) confidence interval width. In other words, sample size based on average width can only have 50% assurance.[17]

Under the assumption of data being normally distributed, a variance function has previously been derived,[3,8,9] with $f(\theta)$ given by

$$0.0099 e^{-\{\Phi^{-1}(\theta)\}^2} [10\{\Phi^{-1}(\theta)\}^2 + 8 + \frac{2\{\Phi^{-1}(\theta)\}^2 + 8}{r}](r+1)$$

where $\Phi^{-1}(\cdot)$ is the inverse of the cumulative distribution function of the standard normal distribution, and $r$ is the sample size ratio of control group to disease group. This function, together with a less frequently used version that additionally involves a standard deviation ratio parameter, tend to provide conservative sample size estimates;[3,8,19] they were originally derived to improve on the variance based on exponential model[1] which can underestimate the sample size.[8] While a conservative estimate is useful in practice to avoid sample size underestimation, being conservative comes from overestimation of the true variance. More efficient (and still valid) sample size planning can be accomplished with a more accurate variance function.

Here we propose a new variance function by combining the delta method and the results in Bonett[27] or Reiser and Guttman[28]:

$$f(\theta) = \frac{1}{2}[\varphi\{\Phi^{-1}(\theta)\}]^2 \left[ \frac{\{\Phi^{-1}(\theta)\}^2}{(1+B^2)^2} \left\{ r + 1 + \frac{(r+1)B^4}{r} \right\} + \frac{2(r+1)}{1+B^2} + \frac{2(r+1)B^2}{r(1+B^2)} \right]$$

where $\Phi^{-1}(\cdot)$ is the inverse of the cumulative distribution function of the standard normal distribution, $\varphi(\cdot)$ is the probability density function of the standard normal distribution, $r$ is the sample size ratio of control group to disease group, and $B$ is the ratio of control participant standard deviation to diseased participant standard deviation. Compared with the conservative option, our variance function needs to additionally specify the value of $B$, but is intended to give more efficient sample size estimates.

Note that $\pi/3$, the reciprocal for the well-known nonparametric statistic relative efficiency,[29] serves as an inflation factor in sample size calculation, given that the nonparametric variance estimation by DeLong et al[30] is commonly used in the stage of data analysis.

**Proposed sample size formula: estimation of a difference between two AUCs**

We now present sample size required for achieving a pre-specified lower bound of the two-sided 95% confidence interval for the difference of two correlated AUCs with assurance probability $1 - \beta$. Let $\theta_1$, $\theta_2$, and $\Delta_0$ denote the pre-specified first AUC, second AUC, and the lower bound, respectively. Define $\theta^* = \frac{\theta_2 - \theta_1 + 1}{2}$ and $\theta_0^* = \frac{\Delta_0 + 1}{2}$. Applying the logit transformation to $\theta^*$ results in constructing confidence interval for a difference between two AUCs ($\Delta$) based on a transformation of $\log\left(\frac{1+\Delta}{1-\Delta}\right)$, which has been applied to case of a difference between two proportions.[31]

After derivations in the Online Supplementary Material, we arrive at the sample size formula below:

$$n = \left\{\frac{z_\beta + z_{\alpha/2}}{\text{logit}(\theta^*) - \text{logit}(\theta_0^*)}\right\}^2 \frac{f(\theta^*)}{(\theta^*)^2(1-\theta^*)^2} \frac{\pi}{3}$$

where

$$f(\theta^*) = \frac{1}{4}\left\{f^{(1)}(\theta_1) + f^{(2)}(\theta_2) - 2\rho\sqrt{f^{(1)}(\theta_1)}\sqrt{f^{(2)}(\theta_2)}\right\},$$

$$f^{(1)}(\theta_1) = \frac{1}{2}[\varphi\{\Phi^{-1}(\theta_1)\}]^2 \left[\frac{\{\Phi^{-1}(\theta_1)\}^2}{(1+B_1^2)^2}\left\{r+1+\frac{(r+1)B_1^4}{r}\right\} + \frac{2(r+1)}{1+B_1^2} + \frac{2(r+1)B_1^2}{r(1+B_1^2)}\right],$$

and

$$f^{(2)}(\theta_2) = \frac{1}{2}[\varphi\{\Phi^{-1}(\theta_2)\}]^2 \left[\frac{\{\Phi^{-1}(\theta_2)\}^2}{(1+B_2^2)^2}\left\{r+1+\frac{(r+1)B_2^4}{r}\right\} + \frac{2(r+1)}{1+B_2^2} + \frac{2(r+1)B_2^2}{r(1+B_2^2)}\right]$$

with $\rho$ being the correlation between two estimated AUCs (from two diagnostic tests), $r$ the sample size ratio of control group to disease group as before, and $B_t$ the standard deviation ratio of control group to disease group for Test $t$ ($t = 1,2$). Since two estimated AUCs are calculated from the same group of participants, $\rho$ is likely to be positive; such correlation, if exists, should be taken into account in statistical procedures.[5]

**Evaluation**

We conducted an extensive series of simulations to evaluate the performance of the proposed sample size formulas. Two criteria used are: i) the empirical assurance probability (in percent), defined as the percentage of the lower confidence limits in 10,000 simulation runs that exceeded the pre-specified lower bound, and ii) the empirical coverage (in percent), defined as the percentage of 95% confidence intervals in 10,000 simulation runs that covered the true value of parameters (i.e., single AUC for one-sample setting and the difference of two AUCs for two-sample setting). Valid and efficient sample size planning is expected to have empirical assurance probability close to the pre-specified assurance probability and empirical coverage close to the nominal 95%. Simulation set-up is described below.

*Configurations:* We applied the proposed formulas to compute required sample size under various configurations: a pre-specified confidence level at 0.95, an assurance probability of 50% or 80%, and a distance of 0.05 or 0.1 between the estimand and the lower bound, denoted by $d$, (i.e., $d = \theta - \theta_0$ in one-sample case with $\theta = 0.7, 0.8$ or $0.9$ and $d = \theta_2 - \theta_1 - \Delta_0$ in two-sample case with $\theta_2 - \theta_1 = 0.2$). We also considered various combinations of the group size ratio $r$ and the standard deviation ratio $B$, each equal to 1 or 2; in two-sample setting we tested a reasonable situation where $B_1$ and $B_2$ are equal and denoted by $B$.

*Data generation process:* Once the sample size was obtained using the formula proposed, we simulated 10,000 data sets each consisting of test values for the disease group from the standard normal distribution and test values for the non-disease group from a normal distribution whose mean and standard deviation were determined by underlying configuration. For two-sample setting, to allow for weak, moderate or strong correlation between two estimated AUCs, we specified a correlation between rating data of two tests as 0.2, 0.5 or 0.8. This correlation, once specified, can be used to obtain the correlation between two estimated AUCs. Specifically, we generated sample of size 5,000 with specified rating data correlation, calculated its empirical correlation between two estimated AUCs, repeated the procedure 5,000 times, and computed the average value. This yielded approximated between-AUC correlation 0.15, 0.42 or 0.71 when $B = 1$ and 0.13, 0.37 or 0.63 otherwise. While such transformation between two correlations was needed in simulations in order to obtain the between-AUC correlation, in practice, users can specify the anticipated between-AUC correlation directly.

*Analyzing simulated data*: For each simulated data set, we constructed the 95% confidence interval using DeLong nonparametric method with the logit transformation applied to $\theta$ for a single AUC and $\theta^*$ for the setting of a difference between two AUCs.

For further illustration of the proposed method, we shall present worked examples based on a prior ROC study[32] that compared multiple tracer kinetic analysis methods in terms of ability to provide good diagnosis of myocardial ischemia.

The simulation study was performed using R Version 3.6.1 software,[33] with the pROC package[12] used to implement DeLong nonparametric variance estimation. An Institutional Review Board approval is not needed because this work did not involve any human participants or animals.

# RESULTS

**Performance of proposed method**

We reported the empirical assurance probability (in percent) and the empirical coverage (in percent) for evaluation. Results in Table 1 suggest for one-sample setting, the empirical assurance probability is close to the nominal levels in virtually all scenarios considered. For example, the formula predicts that if the true AUC = 0.9, a study with a total sample size of 412 with a 1:1 control to disease ratio and common variance would have a lower limit to be above 0.85 with 80% probability, comparable with 83.44% as estimated based on 10,000 simulation runs.

Similar conclusion can be drawn for two-sample setting from results in Table 2. For example, if the two AUCs are 0.9 and 0.7, the formula predicts a total sample of 446 with equal group size and common variance would provide a lower limit to be above 0.15 with 80% assurance, comparable with the estimated value of 82.34% based on 10,000 simulation runs. There are few cases where the estimated assurance probabilities are lower than the nominal level. Specifically, when two tests were strongly correlated, pre-specified assurance probability = 50%, $\theta_2 - \theta_1 - \Delta_0 = 0.1$. This suboptimal performance corresponds to the smallest estimated total sample size of approximately 60. Promising results on assurance probability were observed for sample size as small as 66 in one-sample setting and 92 in two-sample setting. These results showed the proposed sample size formulas performed reasonably well under a wide range of conditions.

The empirical coverage results were quite close to the nominal 95% in all cases. These results demonstrated that combining the DeLong nonparametric variance estimation with the logit transformation produced accurate confidence intervals even when the total sample size is as small as 63 in one-sample setting and 56 in two-sample setting.

**Worked examples**

As an illustration, consider a study conducted by Biglands et al[32] who compared four tracer kinetic analysis methods for diagnosis of myocardial ischemia. The study involved 50 participants, 31 without ischemia and 19 with ischemia. Among all methods, Fermi method gave the highest AUC estimate of 0.92 and the one-compartment model produced the lowest AUC estimate of 0.80, both using myocardial perfusion reserve as the continuous measure in ROC analyses. These two methods were found significantly different with an estimated AUC difference of 0.12 (CI 0.02, 0.21).

Suppose a future study is being planned and similar to the above study. Thus it is reasonable to specify the group size ratio as 1.6 (close to $\frac{31}{19}$), the standard deviation ratio for the Fermi method as 1.1 (close to $\frac{1.02}{0.93}$), and the standard deviation ratio for the one-compartment model as 1.2 (close to $\frac{1.31}{1.11}$), where the standard deviations 1.02, 0.93, 1.31 and 1.11 were reported in Biglands et al.[32]

*Assume the true AUC is 0.92 for the Fermi method. What is the required sample size to achieve a lower bound of 0.8?* With $\theta = 0.92$, $r = 1.6$ and $B = 1.1$, we calculate $f(\theta)$ as $f(0.92) = 0.0679$. With 80% assurance, the required sample size for the disease and control groups is given by

$$n_D = \frac{n}{r+1} = 35.5$$

and

$$n_C = \frac{nr}{r+1} = 56.9,$$

respectively. Since sample size is integer, we require 36 participants with ischemia and 57 without ischemia, adding up to a total of 93 participants. Achieving assurance of 90% requires 125 participants with 48 ischemia cases and 77 controls. When the sample size is 50 as was used in the original study,[32] the corresponding assurance probability is 53%.

*Assume the true AUCs for the one-compartment model and the Fermi method are 0.80 and 0.92, respectively. What is the required sample size to achieve a user-specified lower bound of the AUC difference?* With $\theta_1 = 0.80$, $\theta_2 = 0.92$, $r = 1.6$, $B_1 = 1.2$ and $B_2 = 1.1$, we calculated $f^{(1)}(\theta_1)$ and $f^{(2)}(\theta_2)$ as $f^{(1)}(0.80) = 0.1865$ and $f^{(2)}(0.92) = 0.0679$, respectively. While the correlation between two AUCs is not available from Biglands et al, the authors noted high correlations (≥ 0.88) between tracer kinetic analysis methods in

terms of myocardial blood flow measures. Assuming similar correlations for myocardial perfusion reserve measures, it is reasonable to specify $\rho = 0.8$ and calculate $f(\theta^*)$ as

$$f\left(\frac{0.92 - 0.80 + 1}{2}\right) = \frac{1}{4}(0.1865 + 0.0679 - 2 \times 0.8\sqrt{0.1865}\sqrt{0.0679}) = 0.0186.$$

Assume a lower confidence limit of 0.02 as reported by Biglands et al. Then $\theta_0^* = \frac{0.02+1}{2} = 0.51$. For 80% assurance, $n_D = 23.9$ and $n_C = 38.3$. Thus, the study requires 24 participants with ischemia and 39 without ischemia. For 90% assurance, 85 participants are required, including 33 with ischemia and 52 without.

Fixing 80% assurance, if we wish to reach a larger lower bound at 0.05, 127 participants are required. Finally, ignoring the correlation and specifying $\rho = 0$ leads to a perhaps overly conservative sample size of 434.

**DISCUSSION**

We have proposed simple formulas for computing sample sizes required for achieving pre-specified lower bounds of the confidence intervals for single AUCs and their differences. Our method allows for unequal variances between test values of the disease and non-disease groups and takes into account the discrepancy in variance estimation comparing sample size planning and data analysis phases (parametric versus nonparametric). Simulation results demonstrated satisfactory performance of the proposed formulas in various practical settings in terms of empirical assurance probability and coverage, even when the total sample sizes was as small as 70.

Compared to the proposed variance function $f(\theta)$, the routinely-used variance function in Obuchowski[3,8,9] led to conservative sample size estimates with empirical assurance probability notably higher than the pre-specified assurance probability (**Table 3** vs. **Table 1**). Nevertheless, this existing approach does not require specification of the standard deviation ratio $B$ and thereby is useful when little is known about $B$ and when conservative sample size is not considered a concern. When data distributions other than binormal is appropriate, $f(\theta)$ can be replaced with other existing variance functions such as Equations 6.4 and 6.6 in Zhou et al.[3]

When specifying correlation ($\rho$) in the formula for a difference between two AUCs, a plausible value of the correlation may be obtained with guidance from subject matter knowledge such as prior, similar work in the literature.[30] Pilot data, if available, can also be used to estimate the correlation between the two areas.[8] When information on the correlation between observations of the test (i.e., raw data) is available, that correlation can be used to obtain the correlation between two AUCs.[5] When little is known about either correlation, one conservative approach is to specify $\rho = 0$ given that $\rho \geq 0$ in most practical settings. An even more conservative version is to let $\rho = -1$ but a negative $\rho$ is very unlikely. These two strategies are straightforward to implement but can severely overestimate the sample size when the true correlation is close to 1.

Sometimes the investigator decides to recruit participants by taking a random sample from a large population in which the disease prevalence $p_D$ is known *a priori*. In this case, the group size ratio $r$ can be specified as $\frac{1-p_D}{p_D}$, because $p_D = \frac{n_D}{n_D+n_C} = \frac{1}{1+r}$.

We emphasize that it is always a good idea to investigate sensitivity of sample size estimates under a wide range of values for input parameters for the formulas.

The proposed approach is consistent with the trend towards using confidence intervals for inference.[20,23] In a similar spirit, it is useful to develop sample size formulas for achieving a pre-specified confidence interval width; this direction was previously investigated in reliability studies[17] motivated by inadequate sample size from methods based on expected interval widths.[17,23,34] Extending the work presented here to accommodate partial AUC is another future topic, as we have limited our discussion on area under the entire ROC curve. However, while sometimes it may be clinically meaningful to examine partial AUC over some range of specificity,[35,36] partial area is used less frequently than full area in practice.

## CONCLUSION

Closed-from sample size formulas were proposed to help plan ROC studies focusing on estimation of single AUCs and their differences. This approach achieves a desired chance that the lower confidence limit will be above a pre-specified bound. A free online calculator is developed to implement the formulas.

**DISCLOSURES OF CONFLICTS OF INTEREST AND FUNDING:** none declared.

**ONLINE SUPPLEMENTARY MATERIAL IS AVAILABLE.**

**TABLE 1**. Performance of the proposed sample size formula for estimating a single AUC ($\theta$) with pre-specified lower limit ($\theta_0$) and assurance probability 50% or 80%, under various standard deviation ratio of control group to disease group ($B$) and sample size ratio of control group to disease group ($r$).

| $\theta$ | $\theta_0$ | $B$ | $r$ | 50% assurance | | | 80% assurance | | |
|---|---|---|---|---|---|---|---|---|---|
| | | | | $n$ | ECP | EAP | $n$ | ECP | EAP |
| 0.9 | 0.85 | 1 | 1 | 202 | 95.13 | 51.81 | 412 | 95.08 | 83.44 |
| | | | 2 | 228 | 94.80 | 51.91 | 464 | 94.83 | 84.25 |
| | | 2 | 1 | 224 | 94.84 | 48.56 | 456 | 95.08 | 80.79 |
| | | | 2 | 194 | 94.81 | 49.12 | 393 | 94.94 | 81.23 |
| | 0.80 | 1 | 1 | 66 | 95.15 | 50.68 | 136 | 95.11 | 86.88 |
| | | | 2 | 75 | 94.96 | 52.08 | 152 | 95.18 | 86.77 |
| | | 2 | 1 | 74 | 94.54 | 49.02 | 150 | 94.88 | 84.11 |
| | | | 2 | 63 | 95.25 | 47.52 | 129 | 95.48 | 84.25 |
| 0.8 | 0.75 | 1 | 1 | 352 | 95.23 | 50.96 | 716 | 95.34 | 82.81 |
| | | | 2 | 395 | 94.99 | 51.89 | 806 | 94.82 | 82.85 |
| | | 2 | 1 | 370 | 95.46 | 49.37 | 756 | 95.15 | 80.66 |
| | | | 2 | 326 | 95.14 | 49.74 | 665 | 95.24 | 80.42 |
| | 0.7 | 1 | 1 | 100 | 95.70 | 51.44 | 204 | 95.36 | 84.01 |
| | | | 2 | 113 | 95.17 | 51.65 | 230 | 95.04 | 84.02 |
| | | 2 | 1 | 106 | 95.15 | 48.05 | 216 | 95.47 | 81.34 |
| | | | 2 | 93 | 95.27 | 48.66 | 191 | 95.38 | 82.09 |
| 0.7 | 0.65 | 1 | 1 | 454 | 95.09 | 51.22 | 926 | 94.80 | 81.26 |
| | | | 2 | 510 | 95.06 | 50.60 | 1041 | 95.35 | 80.97 |
| | | 2 | 1 | 464 | 95.47 | 48.10 | 946 | 94.92 | 79.06 |
| | | | 2 | 413 | 95.40 | 48.64 | 843 | 94.50 | 78.63 |
| | 0.6 | 1 | 1 | 122 | 95.30 | 51.23 | 248 | 95.42 | 81.36 |
| | | | 2 | 137 | 95.07 | 50.31 | 279 | 95.54 | 82.07 |
| | | 2 | 1 | 124 | 95.11 | 47.04 | 254 | 95.25 | 79.17 |
| | | | 2 | 111 | 95.10 | 48.23 | 225 | 94.88 | 79.94 |

$n$, estimated sample size; ECP, empirical coverage (in percent), estimated by percentage of times that the 2-sided 95% confidence intervals contain the true value across 10,000 simulated data sets; EAP, empirical assurance probability, estimated by percentage of times that the lower limits of 2-sided 95% confidence intervals being above the pre-specified value across 10,000 simulated data sets.

**TABLE 2.** Performance of the proposed sample size formula for estimating a difference between two AUCs ($\theta_1 = 0.7$ vs $\theta_2 = 0.9$) with pre-specified lower limit ($\Delta_0$) and assurance probability 50% or 80%, under various degrees of correlation between two AUCs and various combinations of standard deviation ratio of control group to disease group ($B$) and sample size ratio of control group to disease group ($r$).

| Correlation | $\Delta_0$ | $B$ | $r$ | 50% assurance | | | 80% assurance | | |
|---|---|---|---|---|---|---|---|---|---|
| | | | | $n$ | ECP | EAP | $n$ | ECP | EAP |
| Strong | 0.15 | 1 | 1 | 218 | 94.93 | 49.36 | 446 | 94.62 | 82.34 |
| | | | 2 | 245 | 94.98 | 48.98 | 501 | 94.90 | 81.65 |
| | | 2 | 1 | 262 | 94.43 | 48.02 | 536 | 95.39 | 80.61 |
| | | | 2 | 234 | 94.58 | 49.08 | 479 | 95.16 | 82.92 |
| | 0.1 | 1 | 1 | 56 | 94.62 | 44.33 | 114 | 94.77 | 83.76 |
| | | | 2 | 63 | 94.25 | 44.96 | 128 | 94.25 | 83.77 |
| | | 2 | 1 | 68 | 94.30 | 46.06 | 136 | 94.30 | 82.27 |
| | | | 2 | 60 | 94.40 | 46.45 | 122 | 94.66 | 83.60 |
| Moderate | 0.15 | 1 | 1 | 360 | 94.61 | 50.85 | 736 | 95.09 | 82.32 |
| | | | 2 | 405 | 95.02 | 49.41 | 828 | 95.18 | 81.84 |
| | | 2 | 1 | 398 | 95.01 | 47.81 | 814 | 94.92 | 79.79 |
| | | | 2 | 354 | 94.95 | 48.58 | 722 | 94.49 | 80.19 |
| | 0.1 | 1 | 1 | 92 | 94.68 | 49.21 | 188 | 94.84 | 82.63 |
| | | | 2 | 104 | 94.81 | 50.26 | 210 | 94.89 | 83.25 |
| | | 2 | 1 | 102 | 94.86 | 47.50 | 208 | 94.75 | 80.72 |
| | | | 2 | 90 | 94.72 | 47.32 | 183 | 95.14 | 81.85 |
| Weak | 0.15 | 1 | 1 | 494 | 95.01 | 50.40 | 1008 | 95.00 | 81.69 |
| | | | 2 | 555 | 95.22 | 50.75 | 1133 | 94.99 | 81.61 |
| | | 2 | 1 | 524 | 94.98 | 48.60 | 1070 | 95.21 | 79.32 |
| | | | 2 | 464 | 94.68 | 49.15 | 945 | 95.28 | 79.82 |
| | 0.1 | 1 | 1 | 126 | 95.25 | 49.77 | 256 | 94.75 | 83.08 |
| | | | 2 | 141 | 94.99 | 50.16 | 288 | 94.55 | 82.53 |
| | | 2 | 1 | 134 | 94.96 | 47.58 | 272 | 94.98 | 80.52 |
| | | | 2 | 119 | 94.63 | 48.11 | 240 | 95.05 | 80.37 |

$n$, estimated sample size; ECP, empirical coverage (in percent), estimated by percentage of times that the 2-sided 95% confidence intervals contain the true value across 10,000 simulated data sets; EAP, empirical assurance probability, estimated by percentage of times that the lower limits of 2-sided 95% confidence intervals being above the pre-specified value across 10,000 simulated data sets.

**TABLE 3** When using the existing variance function in Obuchowski:[8,9] Performance of the proposed sample size formula for estimating a single AUC ($\theta$) with pre-specified lower limit ($\theta_0$) and assurance probability 50% or 80%, under various standard deviation ratio of control group to disease group ($B$) and sample size ratio of control group to disease group ($r$).

| $\theta$ | $\theta_0$ | $B$ | $r$ | 50% assurance | | | 80% assurance | | |
|---|---|---|---|---|---|---|---|---|---|
| | | | | $n$ | ECP | EAP | $n$ | ECP | EAP |
| 0.9 | 0.85 | 1 | 1 | 318 | 95.06 | 73.54 | 650 | 95.37 | 96.51 |
| | | | 2 | 402 | 95.12 | 78.55 | 821 | 94.74 | 97.92 |
| | | 2 | 1 | 318 | 94.72 | 64.57 | 650 | 95.01 | 92.69 |
| | | | 2 | 402 | 95.35 | 82.37 | 821 | 95.01 | 98.85 |
| | 0.80 | 1 | 1 | 104 | 95.29 | 74.62 | 212 | 95.13 | 97.35 |
| | | | 2 | 132 | 95.39 | 80.57 | 267 | 94.65 | 98.43 |
| | | 2 | 1 | 104 | 95.09 | 65.89 | 212 | 95.20 | 94.94 |
| | | | 2 | 132 | 95.09 | 85.48 | 267 | 95.11 | 99.39 |
| 0.8 | 0.75 | 1 | 1 | 454 | 95.34 | 63.01 | 928 | 94.96 | 91.05 |
| | | | 2 | 551 | 94.94 | 65.54 | 1124 | 95.11 | 92.89 |
| | | 2 | 1 | 454 | 95.10 | 57.69 | 928 | 95.19 | 88.22 |
| | | | 2 | 551 | 94.89 | 72.04 | 1124 | 95.14 | 95.81 |
| | 0.7 | 1 | 1 | 130 | 95.59 | 63.53 | 266 | 95.46 | 92.31 |
| | | | 2 | 158 | 94.92 | 67.13 | 321 | 95.33 | 93.93 |
| | | 2 | 1 | 130 | 95.58 | 57.50 | 266 | 95.29 | 89.15 |
| | | | 2 | 158 | 95.15 | 74.30 | 321 | 94.77 | 96.50 |
| 0.7 | 0.65 | 1 | 1 | 510 | 95.17 | 56.43 | 1040 | 94.86 | 85.34 |
| | | | 2 | 594 | 95.23 | 57.99 | 1214 | 94.75 | 86.44 |
| | | 2 | 1 | 510 | 94.97 | 51.61 | 1040 | 94.85 | 81.93 |
| | | | 2 | 594 | 95.47 | 63.54 | 1214 | 94.70 | 91.20 |
| | 0.6 | 1 | 1 | 136 | 95.38 | 55.33 | 278 | 95.09 | 86.16 |
| | | | 2 | 159 | 95.00 | 57.25 | 324 | 95.02 | 87.46 |
| | | 2 | 1 | 136 | 95.46 | 51.59 | 278 | 94.82 | 82.44 |
| | | | 2 | 159 | 95.22 | 63.46 | 324 | 95.15 | 92.33 |

$n$, estimated sample size; ECP, empirical coverage (in percent), estimated by percentage of times that the 2-sided 95% confidence intervals contain the true value across 10,000 simulated data sets; EAP, empirical assurance probability, estimated by percentage of times that the lower limits of 2-sided 95% confidence intervals being above the pre-specified value across 10,000 simulated data sets.

# Supplemental Material for

# "Revisiting sample size planning for receiver operating characteristic studies: a confidence interval approach with precision and assurance"

Di Shu[1,2,3] and Guangyong Zou[4,5]

[1]Department of Biostatistics, Epidemiology and Informatics, University of Pennsylvania Perelman School of Medicine, Philadelphia, Pennsylvania, USA

[2]Department of Pediatrics, University of Pennsylvania Perelman School of Medicine, Philadelphia, Pennsylvania, USA

[3]Center for Pediatric Clinical Effectiveness, Children's Hospital of Philadelphia, Philadelphia, Pennsylvania, USA

[4]Department of Epidemiology and Biostatistics, Schulich School of Medicine & Dentistry, Western University, London, Ontario, Canada

[5]Robarts Research Institute, Western University, London, Ontario, Canada

This supplementary material provides derivations of two proposed sample size formulas.

## Derivations for one-sample setting

Application of the delta method[1,2] with a logit transformation for AUC yields a general sample size formula as

$$n = \left\{\frac{z_\beta + z_{\alpha/2}}{\text{logit}(\theta) - \text{logit}(\theta_0)}\right\}^2 \frac{f(\theta)}{\theta^2(1-\theta)^2} \quad [1]$$

where the logit transformation gives $\text{logit}(\theta) = \log\left(\frac{\theta}{1-\theta}\right)$, $z_\beta$ is the upper $\beta$ quantile of the standard normal distribution, $1 - \alpha$ is the confidence level (conventionally, $\alpha = 0.05$), and $f(\theta)$ is the kernel of the variance for the estimated AUC (factoring out the sample size), that is, $f(\theta)/n = \text{var}(\hat{\theta})$.

To obtain $f(\theta)$, a parametric distribution of data is typically specified so that the variance can be described by a few parameters. Let random samples $X_i$ $(i = 1, \ldots, n_C) \sim N(\mu_C, \sigma_C^2)$ and $Y_j$ $(j = 1, \ldots, n_D) \sim N(\mu_D, \sigma_D^2)$ be normally-distributed test values for participants without disease (control) and for participants with disease, respectively, where $n_D$ denotes the size of the disease group, $n_C$ the size of the control group, and $r$ the sample size ratio of control group to disease group. Define $B$ as the standard deviation ratio of control group to disease group, that is, $B = \sigma_C/\sigma_D$.

As shown by Bamber,[3] the AUC is defined as $P(Y > X) + 0.5 P(Y = X)$ where $P(Y = X) = 0$ for continuous test values. Provided that the test values follow a binormal distribution, the AUC can be written as $\Phi\left(\frac{\eta}{\sqrt{2}}\right)$ where $\eta = \frac{\mu_D - \mu_C}{\sqrt{\frac{\sigma_D^2 + \sigma_C^2}{2}}}$ and $\Phi(\cdot)$ is the cumulative distribution function of the standard normal distribution, leading to an AUC estimator

$$\hat{\theta} = \Phi\left(\frac{\hat{\eta}}{\sqrt{2}}\right)$$

where

$$\hat{\eta} = \frac{\frac{1}{n_D}\sum_{j=1}^{n_D} Y_j - \frac{1}{n_C}\sum_{i=1}^{n_C} X_i}{\sqrt{\frac{\hat{\sigma}_D^2 + \hat{\sigma}_C^2}{2}}}$$

Based on the delta method, we approximate $var(\hat{\theta})$ by $\frac{1}{2}\left\{\varphi\left(\frac{\eta}{\sqrt{2}}\right)\right\}^2 var(\hat{\eta})$ where $\varphi(\cdot)$ is the derivative of $\Phi(\cdot)$ and hence the probability density function of the standard normal distribution, and $var(\hat{\eta})$ is readily approximated by results available in Bonett,[4] which turn out to be equivalent to results in Reiser and Guttman.[5] Specifically,

$$var(\hat{\eta}) \approx \frac{\eta^2}{2(\sigma_D^2 + \sigma_C^2)^2}\left(\frac{\sigma_D^4}{n_D} + \frac{\sigma_C^4}{n_C}\right) + \frac{2\sigma_D^2}{(\sigma_D^2 + \sigma_C^2)n_D} + \frac{2\sigma_C^2}{(\sigma_D^2 + \sigma_C^2)n_C}$$

$$= \frac{\eta^2}{2(1+B^2)^2}\left\{\frac{r+1}{n} + \frac{(r+1)B^4}{rn}\right\} + \frac{2(r+1)}{(1+B^2)n} + \frac{2(r+1)B^2}{(1+B^2)rn}$$

where $n = n_D + n_C$ is the total sample size. Factoring $n$ out and substituting $\eta = \sqrt{2}\Phi^{-1}(\theta)$ yields

$$f(\theta) = \frac{1}{2}[\varphi\{\Phi^{-1}(\theta)\}]^2 \left[\frac{\{\Phi^{-1}(\theta)\}^2}{(1+B^2)^2}\left\{r+1+\frac{(r+1)B^4}{r}\right\} + \frac{2(r+1)}{1+B^2} + \frac{2(r+1)B^2}{r(1+B^2)}\right]$$

While sample size now can be computed by substituting above $f(\theta)$ into equation 1, the resulting sample size may still be inadequate. The reason is that the DeLong nonparametric approach[6] to the analysis of a single AUC or areas under two correlated ROC curves has been commonly-used in practice, with different efficiency compared to a parametric approach based on which the sample size is planned. Ignoring this discrepancy might lead to inadequate sample size, because the nonparametric estimation is slightly less efficient than the parametric countertpart.

To deal with this discrepancy, we further inflate the sample size by $\pi/3$, the reciprocal for the well-known nonparametric statistic relative efficiency,[7] and arrive at the proposed sample size formula:

$$n = \left\{\frac{z_\beta + z_{\alpha/2}}{\text{logit}(\theta) - \text{logit}(\theta_0)}\right\}^2 \times \frac{\pi[\varphi\{\Phi^{-1}(\theta)\}]^2 \left[\frac{\{\Phi^{-1}(\theta)\}^2}{(1+B^2)^2}\left\{r + 1 + \frac{(r+1)B^4}{r}\right\} + \frac{2(r+1)}{1+B^2} + \frac{2(r+1)B^2}{r(1+B^2)}\right]}{6\theta^2(1-\theta)^2} \quad [2]$$

Notably, it can be shown that $(r, B) = (a, b)$ and $(r, B) = (1/a, 1/b)$ produce the same estimated total sample size.

## Derivations for two-sample setting

We begin by introducing the notations. For Test $t$ where $t = 1, 2$, let random samples $X_i^{(t)}$ ($i = 1, \ldots, n_C$) $\sim N(\mu_{Ct}, \sigma_{Ct}^2)$ and $Y_j^{(t)}$ ($j = 1, \ldots, n_D$) $\sim N(\mu_{Dt}, \sigma_{Dt}^2)$ be normally-distributed test values for participants without disease (control) and for participants with disease, respectively. Define $B_t$ as the standard deviation ratio, that is, $B_t = \sigma_{Ct}/\sigma_{Dt}$, for $t = 1, 2$.

We derive the sample size formula through $(\theta_2 - \theta_1 + 1)/2$, which ranges from 0 to 1, in order to leverage formula 1 based on the logit transformation. Noticing that the lower bound of the 95% confidence interval for $(\hat{\theta}_2 - \hat{\theta}_1 + 1)/2$ is $(\Delta_0 + 1)/2$, we define

$$\theta^* = \frac{\theta_2 - \theta_1 + 1}{2}$$

$$\theta_0^* = \frac{\Delta_0 + 1}{2}$$

Note that using $\theta^*$ is equivalent to constructing confidence interval for the difference of AUCs ($\Delta$) based on a transformation $\log\left(\frac{1+\Delta}{1-\Delta}\right)$, a strategy known to perform well for a difference between two proportions.[8]

By equation 1, we obtain a general sample size formula for two-sample case (after applying the efficiency adjustment factor $\pi/3$):

$$n = \left\{\frac{z_\beta + z_{\alpha/2}}{\text{logit}(\theta^*) - \text{logit}(\theta_0^*)}\right\}^2 \frac{\pi f(\theta^*)}{3(\theta^*)^2(1-\theta^*)^2} \quad [3]$$

where $f(\theta^*)$ is the kernel of $var(\theta^*)$ given by

$$var(\theta^*) = \frac{1}{4}\left\{var(\hat{\theta}_1) + var(\hat{\theta}_2) - 2\rho\sqrt{var(\hat{\theta}_1)}\sqrt{var(\hat{\theta}_2)}\right\}.$$

Here $\rho$ is the correlation between $\hat{\theta}_1$ and $\hat{\theta}_2$, the estimated AUCs under two tests.

Factoring $n$ out yields

$$f(\theta^*) = \frac{1}{4}\left\{f^{(1)}(\theta_1) + f^{(2)}(\theta_2) - 2\rho\sqrt{f^{(1)}(\theta_1)}\sqrt{f^{(2)}(\theta_2)}\right\} \quad [4]$$

where

$$f^{(1)}(\theta_1) = \frac{1}{2}[\varphi\{\Phi^{-1}(\theta_1)\}]^2 \left[\frac{\{\Phi^{-1}(\theta_1)\}^2}{(1+B_1^2)^2}\left\{r+1+\frac{(r+1)B_1^4}{r}\right\} + \frac{2(r+1)}{1+B_1^2} + \frac{2(r+1)B_1^2}{r(1+B_1^2)}\right]$$

is the kernel of $var(\hat{\theta}_1)$ and

$$f^{(2)}(\theta_2) = \frac{1}{2}[\varphi\{\Phi^{-1}(\theta_2)\}]^2 \left[\frac{\{\Phi^{-1}(\theta_2)\}^2}{(1+B_2^2)^2}\left\{r+1+\frac{(r+1)B_2^4}{r}\right\} + \frac{2(r+1)}{1+B_2^2} + \frac{2(r+1)B_2^2}{r(1+B_2^2)}\right]$$

is the kernel of $var(\hat{\theta}_2)$.

We arrive at the proposed sample size by substituting $f(\theta^*)$ in equation 4 into equation 3.